\documentclass[aps,prl,twocolumn,superscriptaddress,preprintnumbers,nofootinbib]{revtex4-1}
\pdfoutput=1

\usepackage{graphicx}
\usepackage{bm}
\usepackage{times}
\usepackage{slashed}
\usepackage{amssymb}
\usepackage{color}
\usepackage{slashed}
\usepackage{lipsum}
\usepackage{subfigure}
\usepackage{multirow}
\usepackage{amsmath}
\usepackage{array}
\usepackage{varwidth}
\usepackage{float}

\begin{document}

\title{Observation of Radar Echoes From High-Energy Particle Cascades} 

\author{S. Prohira}%
 \email{prohira.1@osu.edu}
 \affiliation{Center for Cosmology and AstroParticle Physics (CCAPP), The Ohio State University, Columbus OH, USA}
\author{K.D. de Vries}
\affiliation{Vrije  Universiteit  Brussel,  Brussel,  Belgium}

\author{P. Allison}
 \affiliation{Center for Cosmology and AstroParticle Physics (CCAPP), The Ohio State University, Columbus OH, USA}
\author{J. Beatty}
 \affiliation{Center for Cosmology and AstroParticle Physics (CCAPP), The Ohio State University, Columbus OH, USA}
 \author{D. Besson}
 \affiliation{University of Kansas, Lawrence, KS, USA}
 \affiliation{National Research Nuclear University, Moscow Engineering Physics Institute, Moscow, Russia}
 \author{A. Connolly}
  \affiliation{Center for Cosmology and AstroParticle Physics (CCAPP), The Ohio State University, Columbus OH, USA}

  \author{N. van Eijndhoven}
  \affiliation{Vrije  Universiteit  Brussel,  Brussel,  Belgium}

 \author{C. Hast}
 \affiliation{SLAC National Accelerator Laboratory, Menlo Park, CA, USA}
 \author{C.-Y Kuo}
 \affiliation{National Taiwan University, Taipei, Taiwan}
 \author{U.A. Latif}
 \affiliation{University of Kansas, Lawrence, KS, USA}
 \author{T. Meures}
 \affiliation{University of Wisconsin-Madison, Madison, WI, USA}
 \author{J. Nam}
  \affiliation{National Taiwan University, Taipei, Taiwan}
\author{A. Nozdrina}
\affiliation{University of Kansas, Lawrence, KS, USA}
\author{J.P. Ralston}
\affiliation{University of Kansas, Lawrence, KS, USA}
\author{Z. Riesen}
\affiliation{California Polytechnic State University, San Luis Obispo, CA, USA}

\author{C. Sbrocco}
 \affiliation{Center for Cosmology and AstroParticle Physics (CCAPP), The Ohio State University, Columbus OH, USA}

\author{J. Torres}
 \affiliation{Center for Cosmology and AstroParticle Physics (CCAPP), The Ohio State University, Columbus OH, USA}

\author{S. Wissel}
\affiliation{California Polytechnic State University, San Luis Obispo, CA, USA}

\begin{abstract}
  We report the observation of radar echoes  from the ionization trails of high-energy particle cascades. These data were taken at the SLAC National Accelerator Laboratory, where the full electron beam ($\sim$10$^9$ e$^-$ at $\sim$10\,GeV/e$^-$) was directed into a plastic target to simulate an ultra high-energy neutrino interaction. This target was interrogated with radio waves, and coherent radio reflections from the cascades were detected, with properties consistent with theoretical expectations. This is the first definitive observation of radar echoes from high-energy particle cascades, which may lead to a viable neutrino detection technology for energies $\gtrsim 10^{16}$\,eV. 
\end{abstract}

\maketitle

{\bf Introduction.---}
Ultra high energy (UHE; $\gtrsim 10^{16}$ eV) astrophysical neutrinos offer great discovery potential.  They would probe the accelerators of UHE cosmic rays, which are detected up to $\sim$$10^{20}$ eV. Unlike cosmic rays, which are downscattered on the cosmic microwave background and also deflected in magnetic fields, detected neutrinos will point back to their sources. UHE neutrino-nucleon interactions probe center-of-mass energies above the energy scale of colliders, allowing sensitive tests of new physics. To fully exploit the scientific potential of UHE neutrinos, we ultimately need an observatory with sufficient exposure to collect high statistics even in pessimistic flux scenarios.  

When UHE neutrinos interact in matter, they produce a relativistic cascade of particles, as well as a trail of non-relativistic electrons and nuclei produced through the energy loss of the relativistic particles.  The time-integrated cascade profile is a ellipsoid of length $\sim 10$ m and radius $\sim 0.1$\,m, and nearly all of the primary interaction energy goes into ionization of the medium.

The incoherent optical Cherenkov emission from individual cascade electrons and positrons can be detected in TeV--PeV detectors like IceCube~\cite{icecube} and similar experiments~\cite{antares, baikal, km3net}.   However, the optical portion of the proposed successor IceCube-Gen2~\cite{gen2} is too small to be an adequate UHE observatory, due to the steeply falling neutrino spectrum.  Therefore, there are several proposed and implemented methods to detect these cascades from UHE neutrinos.  First, the coherent radio-frequency Cherenkov emission from a net charge asymmetry in the cascade (the Askaryan effect~\cite{askaryan_orig}) has been observed in the lab~\cite{askaryan_detect}, and is the focus of a variety of past~\cite{rice}, present~\cite{ara, anita, arianna}, and proposed~\cite{pueo, rno} experiments. Radio methods can instrument large volumes more sparsely than optical detectors due to the transparency of radio in ice~\cite{zas_halzen_radio, radio_ice_orig, barwick} making the construction of a large detector more cost-effective.  Second, a $\tau$ neutrino, interacting in the earth, can produce a $\tau$ lepton---carrying much of the primary $\nu_{\tau}$ energy---that exits the earth and decays in air, producing a cascade. A current is induced in this cascade as it moves relativistically through the earth's geomagnetic field, leading to coherent radio emission~\cite{scholten_gm, codalema_geomagnetic, lopes, aera_polarization, t510} that might be detected by proposed experiments~\cite{grand, beacon, taroge_m}. Third, the optical Cherenkov light from such in-air decays can be detected by balloon-or satellite-borne experiments~\cite{euso_spb,poemma}.  All of these methods have potential for discovery at very high energies. However, they all have limited sensitivity at the lower end of the UHE range, between 10-100\,PeV, just above the reach of optical Cherenkov detectors like IceCube. 

Finally, it has been proposed that cascades can be detected by radar reflections off the ionization trail left in their wake. This technique shows promising projected sensitivity~\cite{krijnkaelthomas, radioscatter}, and is the only technique forecasted to have peak sensitivity in the 10-100\,PeV range, with the potential to close the gap between optical Cherenkov detectors and the high energy technologies listed above. To that end, several recent experimental efforts~\cite{chiba, chiba2, krijn_els} have made incremental progress toward the detection of a radar echo from a cascade in a dense medium.

 In this letter, we present the first definitive observation of a radar echo from a particle cascade. This observation was made by experiment T576 at the SLAC National Accelerator Laboratory, where their electron beam was used to produce a particle cascade with a density equivalent to that of a $\sim$10$^{19}$\,eV neutrino interaction in ice, and with a similar shower profile. A transmitting antenna (TX) broadcast continuous-wave (CW) radio into the target, and several receiving antennas (RX) monitored the target for a radar reflection. We report on the observation of a signal consistent with theoretical predictions. Below, we detail the experiment, analysis technique, and results.

\begin{figure}[t]
\centering
  \includegraphics[width=0.48\textwidth]{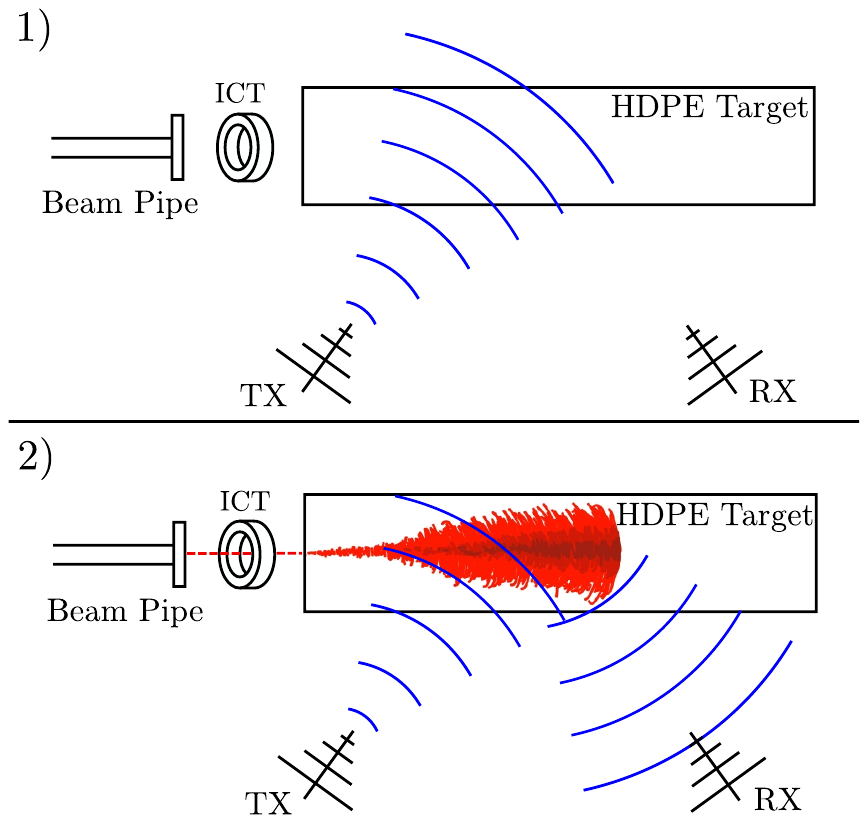}
  \caption{Cartoon of the radar method and T576. 1) A transmitter (TX) illuminates a target. 2) The beam passes through the integrating current toroid (ICT, to monitor beam charge and align the data during analysis) and creates a cascade inside the target, leaving behind an ionization cloud. The transmitted signal is reflected to a monitoring receiver (RX). Not to scale.}
  \label{setup}
\end{figure}

{\bf Experimental setup and data collection.---}
The experiment, depicted in Fig.~\ref{setup}, took place at End Station A at SLAC, a large, open hall with a rich history of discovery. Designated T576, the experiment had two runs during 2018, one in May after which a suggestion of a reflection was reported~\cite{t576_run1}, and a second run in October, which is the focus of the present article. We broadcast CW radio at a range of frequencies between 1 and 2.1 GHz and a range of amplitudes, using a signal generator, 50~W power amplifier, and transmitting antenna (TX) toward a target of high-density polyethylene (HDPE), into which the electron beam was directed. Receiving antennas (RX) were also directed at this target to measure the radar reflection. The data presented in this article were captured by a Tektronix 4 channel, 20~GS/s oscilloscope.

Two different types of antennas were used in this analysis. One was a Vivaldi-style, ultra-wide-band antenna (0.6--6\,GHz) with a measured forward gain of $+$12\,dBi at 2\,GHz, and the other was a custom-built 0.9--4\,GHz log-periodic dipole antenna (LPDA). The LPDA was used in conjunction with a parabolic dish reflector, with a measured forward gain of $+$18\,dBi at 2\,GHz. Surrounding the beam pipe exit was an integrating current toroid (ICT), which gave a precise measurement of the charge in each bunch, and provided a very stable reference point for post-run alignment of the dataset.

The data taking was separated into sub-runs consisting of 100--500 events. Between sub-runs, certain parameters (TX frequency, TX amplitude, TX position and RX position) were varied. Runs in which data was taken for analysis are called signal-runs. Other sub-runs were reserved for taking background data, which will be discussed in the data analysis section, and are called background-runs. The run lasted 8 days, with over 4 full days of beam time acquired in 12-hour increments.

{\bf Expectations.---}The radar method had been suggested for cosmic-ray initiated extensive-air-shower (EAS) detection in the atmosphere as early as 1940~\cite{blackett, early_radar}, with further development in the 1960s~\cite{tokyo_large_air_shower}, followed by stagnation, and then renewed interest in the early 2000s~\cite{gorham, stasielak, bakunov}. Recent experimental searches from terrestrial radar systems~\cite{mu_radar_1} and a dedicated experiment, Telescope Array RAdar (TARA),~\cite{tara, tara_limit} reported no signal due to collisional losses---which limit the efficiency of the scattering---and insufficient ionization density in air. Short free-electron lifetimes ($\tau\sim$\,ns) in air at EAS altitudes cause the ionization to vanish before a sufficient density to reflect incident RF can be achieved. Cascades in ice or other dense media do not suffer from this problem.

The theory for radar is well-established, and models of radar detection of cascades in dense media have evolved to maturity in recent years. Whether built up from a macroscopic~\cite{krijnkaelthomas} or first-principles~\cite{radioscatter} viewpoint, the properties of a reflection are well-defined, and subject to several properties of the material in which the cascade happens. The maximum density of the ionization is directly proportional to the density of the medium. Another critical parameter is the mean ionization lifetime of the material. This lifetime $\tau$ dictates the longitudinal extent of the ionization deposit, and thus the overall length scale of the reflector. For ice, $\tau$ ranges from $\cal{O}$($1-10$\,ns) and is strongly dependent upon the temperature of the ice~\cite{ice_conductivity_1}. For HDPE, the lifetimes are comparable to those of cold polar ice~\cite{hdpe_conductivity}.

For a given transmitter and receiver, the spectral content of the reflected signal is a function of $\tau$ and the cascade geometry. For a compact cascade, as was the case for T576, any lifetime exceeding 1 ns would produce a significant radar reflection at the transmitted frequency. (In nature, an UHE cascade of similar density would be longer by a factor of $\sim$few in ice, which is expected to cause an effective Doppler shift depending upon the radar geometry.) We transmitted at a peak power of 50\,W, with no amplification on our receivers. The expected signal for T576 was a radar return of a few ns in duration, at the transmitter frequency, at a level of a few mV.

{\bf Data analysis.---} The data analysis for T576 was challenging because of the high-amplitude backgrounds. When a charge bunch such as the SLAC beam traverses media with differing indices of refraction, or effective indices of refraction, transition radiation of various forms~\cite{tr, sa, measured_tr} is produced. These signals---which would not be present in nature\footnote{Except for the case of a cascade crossing the air/ice boundary, either an in-ice neutrino cascade breaking out into the air, or an in-air cosmic ray cascade breaking into the ice. Sensitivity to such events is subject to the orientation of transmit and receive antennas, and will be explored.}---exceeded our expected radar signal by a factor of 10-100 in amplitude. We call the total RF background caused by the beam `beam splash' owing to its overall messy character. Fortunately, the beam splash was quite stable, and therefore able to be characterized and filtered using a sensitive matrix-decomposition technique, detailed in~\cite{svd_filter} and based on~\cite{bean_ralston}, that we call singular-value-decomposition filtration, or SVD-filtration.

There are four nominal components to the signal-run data: CW, beam splash, noise, and signal (a radar reflection). The background-run data contains only beam-splash and thermal noise. Assuming that the response of our system is linear for the range of signals received (which we confirmed subsequently in the lab with independent measurements, discussed below) then the total background to our signal can be formed by a linear combination of CW, beam splash, and thermal noise. We call this linear combination  `null data.' To build the null data, we added pre-signal-region CW from signal-run data to signal-region beam splash in background-run data.

SVD-filtration identifies and removes patterns. Patterns are features in the data that are found in multiple individual measurements, such as the beam splash and CW. The SVD-filtration characterizes these patterns within a set of carefully aligned null data, producing a filter basis. Then a filter is produced for each individual event by expanding it in the filter basis. After applying this filter, the only thing remaining in the event should be random, featureless background noise, and any putative signal present in the real data.

The filtration process was a blind procedure, having been tuned on a number of sub-runs comprising $<$10\% of the data. In addition to the null data produced to build the SVD-filter basis, a null event was constructed for every real event in the full dataset, to serve as the null hypothesis. An SVD-filter was constructed for each signal-run according to its associated background-run, and both datasets (real and null) were filtered using the same SVD-filter basis. The resultant filtered data was then analyzed for excess.

\begin{figure}[t]
  \centering
  \includegraphics[width=0.48\textwidth]{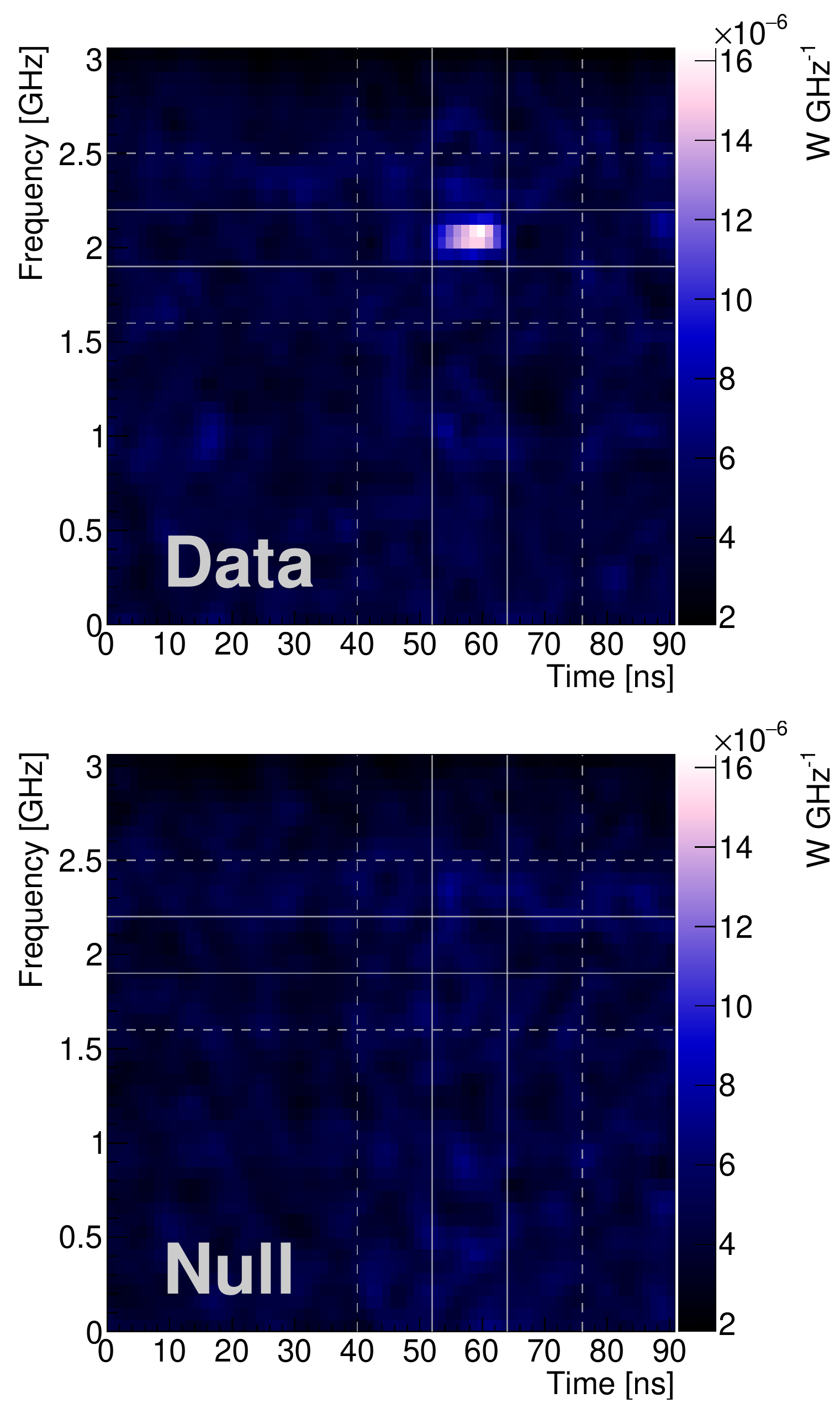}
  \caption{(Top) Time-versus-frequency (spectrogram) representation of the observed signal in data. This is the average of 200 events in a single signal-run. (Bottom) The same representation for the associated null data set for this sub-run. In both plots, the cross-hairs indicate the signal and sideband regions, used to calculate the significance as described in the main text.}
  \label{spectrogram}
\end{figure}

\begin{figure*}[t]
  \centering
  \includegraphics[width=0.98\textwidth]{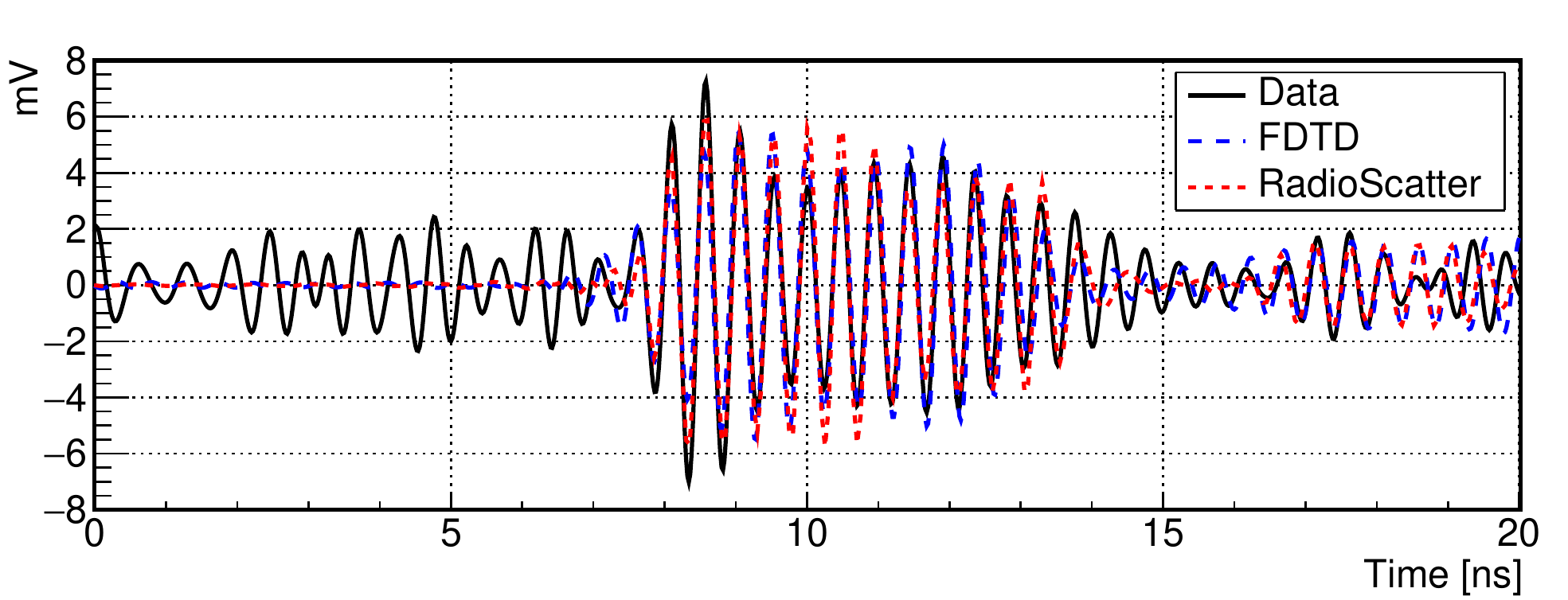}
    \caption{An example time-domain average of the highest-SNR reflections from a signal-run (solid black), compared to the output of an FDTD simulation for the same signal-run (dashed red), and a RadioScatter simulation for the same signal-run (dashed cyan). The plasma lifetime for the simulation is 3\,ns.}
  \label{time_domain}
\end{figure*}

\begin{figure}[h]
  \centering
  \includegraphics[width=0.48\textwidth]{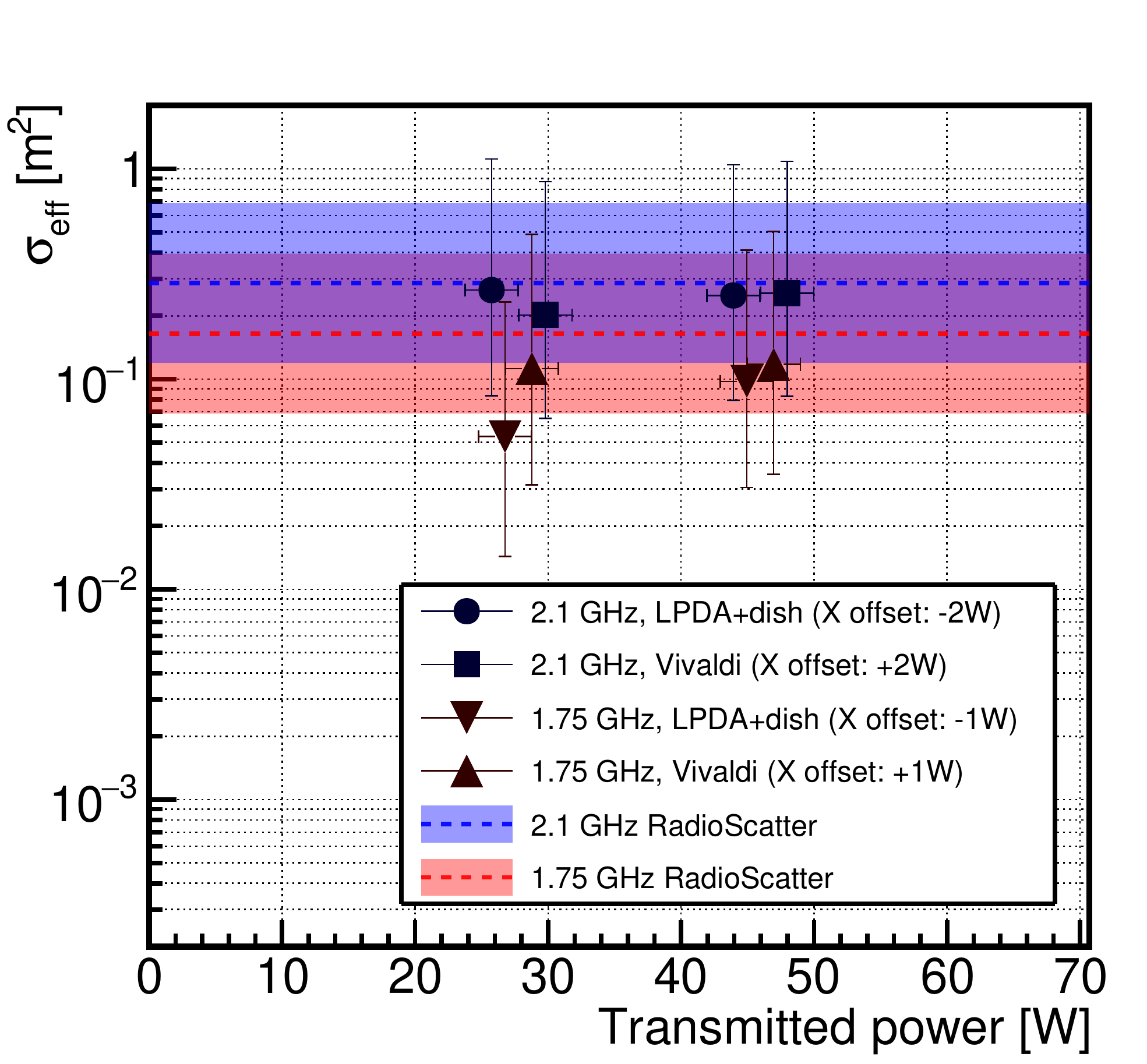}
  \caption{The effective scattering cross section, $\sigma_{\textrm{eff}}$, as a function of transmitter output power, for various receiving antennas (LPDA+dish and Vivaldi) and 2 different frequencies. Errors are statistical and systematic, and dominated by the latter. Each set of 4 points has been clustered around the true X axis value, for clarity, with the offset indicated in the figure legend. The solid bands are $\sigma_{\textrm{eff}}$ from RadioScatter including statistical and systematic errors.}
  \label{rcs}
\end{figure}

{\bf Results.---}After filtration, the dataset was further processed using a method devised during the run-1 analysis. To investigate both the time and spectral content of the signal, a time-versus-frequency spectrogram was generated for each filtered event in a signal-run, and these spectrograms were averaged. The result of such a process is shown in Figure~\ref{spectrogram}, where a clear excess is visible in the real data---and not in the null data---at the transmitter frequency of 2.1\,GHz with a duration of a few ns. A similar excess was observed at many different transmit frequencies, antenna positions, and in different antennas, but no excess is observed at the same time and frequency point in the null data. Signal and sideband regions are indicated by the solid and dashed lines respectively, used in significance calculations defined below.

The highest amplitude signal was expected and received during runs with a horizontally-polarized, high-gain antenna at the specular angle, where the resultant (SVD-filtered) signal was large enough to extract a time-domain waveform through careful alignment and averaging. The alignment was performed so that the events could shift by no more than a fraction of a transmit period, and they were averaged. A resultant time-domain average is shown in Figure~\ref{time_domain}, where only events that had high enough SNR for reliable cross-correlation are used in order to facilitate qualitative comparison to simulation. Also in Figure~\ref{time_domain} is a comparison to an FDTD simulation of the same signal-run (including models of the actual antennas used, the same ionization profile, and the same target material), and a comparison to the RadioScatter simulation code~\cite{radioscatter_github}, which is particle-level and runs within GEANT4~\cite{geant}. The simulations have been scaled by a few percent to allow comparison of the waveform shapes and aligned in time with the data. The plasma lifetime is set to $\tau=3$\,ns in the simulation.

Several checks were performed to establish that the observed signal has properties consistent with a radar scatter. The first and most obvious is the observation that the signal scales with the transmitter output power. This is shown in Figure~\ref{rcs}, where we plot the effective scattering cross section, $\sigma_{\textrm{eff}}$, as a function of transmitter power. This expression, discussed in~\cite{radioscatter}, is a measure of the effective `size' of the reflecting region, should have a weak dependence on frequency at these energies, and should be constant with transmit power. All of these attributes are observed for the signal, which is shown in comparison to RadioScatter simulation (solid bands, including systematic error of HDPE collisional frequency, which is ionization energy dependent~\cite{nist_cross_section}). The errors in the measurement of $\sigma_{\textrm{eff}}$ include statistical and systematic uncertainties, and the systematic errors are tabulated in Table~\ref{errors} along with the dependence of each error. Some errors affect the overall level of all received signal amplitudes (globally dependent) while others would, for example, introduce systematic offsets between antennas (antenna-to-antenna dependent). 
Trending of the signal with TX or RX baselines was not observed, owing to the fact that our antennas were not in the diffractive far field. This non-observation of such trends was verified by FDTD simulations.

\begin{table}
\begin{tabular}{ |l|l|l| }
 \hline
  Systematic & Est. error (dB) & dependence  \\
  \hline\hline
  
  Room effects & 3 & A,F,G\\
  \hline
  Antenna gain/orientation & 1 & A,F\\
  \hline
  Cable loss measurement & 1 & P\\
  \hline
  TX output power & 2 & P,G\\
  \hline
\end{tabular}
\caption{Sources of systematic error (in dB of received power), and their associated estimated errors, used in Figure~\ref{rcs}. Indicated in the right column is the dependence of the individual systematic on the data, either antenna-to-antenna dependent (A), frequency dependent (F), power dependent (P) or globally dependent (G).}
\label{errors}
\end{table}

Because the signal is so small relative to the beam splash, and the null hypothesis relies on a linear combination of background components, an obvious concern is a non-linearity in the overall system. After the run, a series of tests were performed in which CW at the same frequency and amplitude as T576 was amplified and broadcast via a Vivaldi antenna, and another Vivaldi, connected to an oscilloscope, was set up as a receiver. A high-voltage pulse with similar spectral content to the beam splash was broadcast simultaneously. The same analysis technique explained here, involving construction of null data and SVD-filtration, was performed on these data, and no excess was observed. 

To establish a significance against a random fluctuation of the background, we generated $N=10^7$ sets of 100 null events via bootstrapping, made an average spectrogram (like in Figure~\ref{spectrogram}) for each set, and evaluated a test statistic of the sideband-subtracted power excess in the signal region. The signal region, tuned on a discarded subset of the data, is outlined in Figure~\ref{spectrogram}. The value of the test statistic ($\mu$W ns) in the null data is TS$_{\textrm{null}}=2.20^{+6.56}_{-6.20}$. The value of the test statistic in the measured data is TS$_{\textrm{data}}=61.2^{+7.40}_{-6.58}$, well in excess of the 5\,$\sigma$ quantile.

{\bf Conclusions.---} In this letter, we have presented the observation of a radar reflection from a particle-shower induced cascade in a dense material. We have shown that the signal is in good agreement with theoretical expectations, and has a negligible probability of being a background fluctuation. This detection has promising implications for UHE neutrino detection, particularly in the 10-100\,PeV range.
\bigskip

{\bf Acknowledgments.---}We thank the personnel at SLAC for providing us with excellent, stable beam and a safe, productive work environment. We also thank David Saltzberg for invaluable assistance during both runs of T576, and John Beacom for comments and revisions of early drafts of this letter. This work was performed in part under SLAC DOE Contract DE-AC02-76SF00515. SP was partially supported by a US Department of Energy Office  of  Science  Graduate  Student  Research  (SCGSR)  award, administered  by  the  Oak  Ridge  Institute  for  Science  and  Education  for  the  DOE  under contract DE-SC-0014664. Computing resources were provided by the Ohio Supercomputer Center. KDdV was supported in part by  the  Flemish  Foundation  for Scientific  Research  FWO-12L3715N, and  the  European  Research  Council  under  the EU-ropean Unions Horizon 2020 research and innovation programme (grant agreement No 805486). JN and CYK were supported by the Vanguard program from the Taiwan Ministry of Science and Technology.
 
\bibliography{/home/natas/Documents/physics/tex/bib}

\end{document}